\documentclass[structabstract]{aa}   
\usepackage{graphicx}
\usepackage{txfonts}
\begin{document}
 \title{The debris disk host star HD\,61005: \\ a member of the Argus Association?
\thanks{Based on observations collected at La Silla and Paranal Observatory, ESO (Chile). 
Programs 072.A-9006(A), 074.A-9020(A),  076.A-9013(A). 
Based on the All Sky Automated Survey (ASAS) photometric data.}}

  \author{S. Desidera  \inst{1}
          \and
          E. Covino  \inst{2}
          \and
          S. Messina \inst{3}
          \and
          V. D'Orazi \inst{1}
          \and
          J.M. Alcal\'a  \inst{2}
          \and
          E. Brugaletta  \inst{3,4}
          \and
          J. Carson \inst{5}
          \and
          A.C. Lanzafame \inst{3,4}
          \and
          R. Launhardt  \inst{6}
          }

  \institute{INAF-Osservatorio Astronomico di Padova, 
             Vicolo dell'Osservatorio 5,  35122 Padova, Italy;
          \and
         INAF-Osservatorio Astronomico di Napoli, Salita Moiariello 16, 
         80131, Napoli, Italy
           \and 
          INAF-Osservatorio Astrofisico di Catania, 
          Via S. Sofia 78, 95123 Catania, Italy
           \and
          Universit\'a di Catania, Dipartimento di Fisica e Astronomia, 
           Via S. Sofia 78, 95123 Catania, Italy
           \and
          College of Charleston, Department of Physics \& Astronomy, USA
           \and
            Max Plank Institute for Astronomy, Heidelberg, Germany}

\date{}

\abstract
{HD\,61005 is a nearby young solar type star that shows a large infrared excess due to a debris disk. The disk  has
been recently imaged from ground and space, with indications of several components. 
Some characteristics of the disk suggest the presence of planetary companions around the star, that
remain undetected in deep adaptive optics imaging.}
{For a better understanding of the system we aim to refine the determination of the stellar parameters, 
 with emphasis on  the stellar age and system orientation.}
{We used ASAS and Hipparcos photometry and FEROS spectra to determine the rotation period, radial and rotational velocity,
chromospheric emission, effective temperature, and chemical composition.}
{We find no indication of any misalignment between the star rotation axis and the disk. The standard age calibrations
applied to several indicators yield an age close to that of the Pleiades (120 Myr); however the kinematic properties
strongly support its membership in the younger (40 Myr) Argus association, which also includes the IC 2391 open cluster.
Detailed comparison of the properties of HD\,61005 and IC 2391 members shows that the characteristics of HD\,61005 are compatible
with membership to the Argus association, once its rather slow rotation is taken into account, because lithium and other age indicators
are somewhat correlated with stellar rotation at a fixed age. We also identify systematic differences between the field and cluster
population of the Argus association, which are probably  selection effects, so we suggest that additional members with slower
rotation and lower activity level are waiting to be identified.}
{} 

\maketitle

\section{Introduction}
\label{s:intro}

\object{HD\,61005} is a nearby ($\approx 35$\,pc) solar type star that hosts a debris disk discovered 
with Spitzer (Meyer et al.~\cite{meyer06}). The infrared excess with respect to the
stellar photosphere at $24~ \mu m$ is the largest that any star shows in the FEPS sample 
(Meyer et al.~\cite{meyer08}).
The disk was resolved with HST (Hines et al.~\cite{hines07}; 
Maness et al.~\cite{maness09}) and VLT (Buenzli et al.~\cite{buenzli10}) observations; the latter were obtained as part of the NaCo Large Program for Giant 
Planet Imaging (ESO program 184.C-0567).
These studies revealed two components, an external one shaped
by the interactions with the interstellar medium and a ring seen
nearly edge-on. The ring has a semimajor axis of $61.25\pm0.85$ AU and an eccentricity of
$0.045\pm0.015$, and its center is characterized by an offset from the star of $2.75\pm0.85$ AU 
(Buenzli et al.~\cite{buenzli10}).
Furthermore, Fitzgerald et al.~(\cite{fitzgerald10}) report preliminary evidence
of a second, inner belt from  $10~\mu m$ observations, 
whose projection might be misaligned by about $20^{\circ}$ with respect to that of the ring.
These features might be explained by the presence of planetary companion(s).
Indeed similar characteristics in the debris disks of the early type stars \object{$\beta$ Pic}
and \object{Fomalhaut} were found to be accompanied by giant planets (Kalas et al.~\cite{kalas08}; 
Lagrange et al.~\cite{lagrange10} and references therein).

The interpretation of the direct imaging results in terms of planetary mass limits
(Buenzli et al.~\cite{buenzli10}) and of possible future planetary detections
heavily relies on the derived stellar age.
Hines et al.~(\cite{hines07}) estimated the age of HD\,61005 to be $90\pm40$ Myr,
while Weise et al.~(\cite{weise10}) and Roccatagliata et al.~(\cite{roccatagliata09}) report 
30 and 135 Myr, respectively.
Considering the special characteristics of this system, the uncertainty in
the current age determination, and the availability of new observational data, 
we performed a new study of the properties of the
star, with the goal of  improving the stellar age determination.
We also intend to derive the orientation of the star rotation axis and compare it with 
the edge-on orientation of the debris ring ($i=84.3\pm1^{\circ}$, Buenzli et al.~\cite{buenzli10}).
This is of special relevance considering the high frequency
of close-in planets whose normals to the orbital plane are found to be misaligned
with respect to the stellar rotation axis (Triaud et al.~\cite{triaud10}; Winn  et al.~\cite{winn10}). 
Recently, Lai et al.~(\cite{lai10}) have proposed that the interaction between the star and
the disk might produce misaligned and even retrograde configurations.
Systems with resolved disks and late spectral type that allow determination of the rotation period
through photometric monitoring, such as HD\,61005, are suitable targets for testing this hypothesis
(Watson et al.~\cite{watson10}).
An alternative mechanism that produces misaligned systems that might be at work for HD\,61005 
is to add material from the interstellar medium onto a circumstellar disk, which
alters the inclination of the disk (Moeckel \& Throop \cite{moeckel09}).

Our paper is structured as follows. 
In Sect.~\ref{s:prot} we describe the photometric data and the determination of the 
rotation period; 
in Sect.~\ref{s:spec} we present the spectroscopic data and the data reduction procedures, 
and we describe the spectroscopic analysis we performed to derive
radial and projected rotation velocity, chromospheric emission, effective temperature, and
chemical composition;
in Sect.~\ref{s:param}, using the results of the previous sections as well as additional literature data,
we discuss system properties such as orientation, kinematics, and binarity; 
in Sect.~\ref{s:age} we derive the stellar age and discuss the membership to the Argus association, and 
in Sect.~\ref{s:conclusion} we summarize our conclusions.

\section{Rotation period}
\label{s:prot}


We retrieved photometric time series of HD~61005 from both the ASAS All Sky Automatic Survey archive 
(Pojmanski \cite{pojmanski02})
(data collected from November 2000 to November 2009) and the Hipparcos and Tycho epoch photometry 
archives (data collected from November 1989 to February 1993).

For Hipparcos and Tycho photometry, we selected 240 and 183 measurements, 
respectively, quoted as quality A or B. After removing very few
outliers, we were left with 236 and 180 measurements, respectively.
From Hipparcos data we derived a mean magnitude of V=8.22 and a standard deviation $rms$=0.024 mag, 
whereas the data precision was $\sigma$=0.015 mag. From Tycho  data we derived a mean magnitude V=8.21 
and a standard deviation $rms$=0.13 mag. However, the data precision was a much poorer $\sigma$=0.11 mag.
The Hipparcos and Tycho magnitudes were transformed into the Johnson standard system according to the 
equations in Perryman et al.~(\cite{perryman97}).

For ASAS photometry, we retrieved  617 useful measurements (quality A-C). 
Among five different apertures,  we found that a 4-pixel ($\sim$30''
radius) aperture photometry extraction provided the highest photometric accuracy. 
After removing outliers  (by applying a 3$\sigma$ threshold) we 
were left with 588 magnitude values useful for subsequent  period search. The mean magnitude was 
V=8.22 mag  with a dispersion of $rms$=0.028 mag and an average photometric precision of
$\sigma$=0.03 mag. The typical time sampling of ASAS observations was one observation every two to three
days.

\subsection{Search for rotation period}

A visual inspection of the HD~61005 photometric time series clearly shows this star 
is variable. Based on its late spectral type, we expect this variability to arise 
from brightness inhomogeneities (starspots) that are carried in and out of view by the star's rotation, unevenly distributed across the stellar 
photosphere.

To search for significant periodicities in the HD~61005 ASAS time series coming from stellar rotation, 
we sectioned the complete series into 14 time intervals never exceeding about two to three months.
This approach has the twofold advantage of minimizing possible phase shifts introduced by the growth and decay 
of active regions, and of allowing us to reveal possible temporal variations in the rotation period 
that may be related to surface differential rotation. 

The Lomb-Scargle periodogram (Scargle \cite{scargle82}) was computed on each segment, as well on the complete time series.
In the latter case, provided that the light curve remains quite stable 
(as  proved to be the case of HD~61005), 
we can detect the rotation period with a much higher precision and confidence level.
A more detailed description of the data preparation and period search (determination of false alarm 
probability and period uncertainty) can be found in Messina et al.~(\cite{messina10}).

\subsection{Results}

We found that  HD~61005 has a rotation  period of $P=5.04\pm0.04$ days. The same rotation period 
is found in the ASAS, Hipparcos, and Tycho photometry. 
In Table \ref{t:prot} we list the results of our period search. We list the rotation period, its uncertainty, the 
peak-to-peak light curve amplitude, the photometric precision, the JD time interval in which the period 
search was performed, 
the number of 
measurements, the normalized peak power in the Lomb-Scargle periodogram (at 99\% confidence level, 
as computed from simulation), and the normalized
power derived from real data, the brightest magnitude, the source of photometry, and the results derived from 
the time series without sectioning.

\begin{table*}[h]
\caption{Determination of rotational period on various datasets} 
\label{t:prot}
\begin{center}       
\begin{tabular}{cccccccccl} 
\hline
Period &  $\Delta P$  &   $\Delta V$  &    Acccur.  &   JD range   &    \# of  &   PN @99\%  &  PN &  Vmin & Source \\
 (d)   &     (d)      &     (mag)     &     (mag)   &  (2440000.0)&    meas. &   conf. level &     &  (mag)& \\

\hline
5.04   &   0.04  &   0.05  &  0.033 &  11869.8-15137.8  & 588  &   8.03  &   13.1  & 8.22 &  ASAS    full JD range\\
5.05   &   0.15  &   0.03  &  0.034 &  12684.8-12746.8  &  29  &   5.71  &    6.9  & 8.22 &  ASAS    \\
5.08   &   0.15  &   0.04  &  0.034 &  13418.7-13479.7  &  25  &   5.76  &    6.0  & 8.20 &  ASAS  \\
5.09   &   0.15  &   0.05  &  0.033 &  14152.7-14213.7  &  22  &   5.29  &    7.7  & 8.19 &  ASAS  \\
5.04   &   0.20  &   0.05  &  0.015 &   7857.6-9030.1   &  97  &   5.56  &   10.9  & 8.22 &  Hipparcos full JD range\\
5.02   &   0.80  &   0.05  &  0.014 &   8043.8-8254.7   &  25  &   5.54  &    7.1  & 8.21 &  Hipparcos  \\
5.07   &   0.05  &   0.22  &  0.108 &   7857.6-9030.1   & 102  &   3.72  &    6.6  & 8.21 &  Tycho    full JD range\\
\hline
\end{tabular}
\end{center}
\end{table*} 

All the rotation period determinations are consistent with each other within the computed errors. 
The photometric  period variations allow derivation of a lower limit to the amplitude of the 
surface differential rotation (see  Messina \& Guinan \cite{messina03}). In the case of HD~61005, 
these variations are less than 1\%, whereas the  average period uncertainty  is $\Delta$P = 3\%.  
Therefore, based on the available photometry we can exclude the presence of significant 
($\ge 1\%$ at 3$\sigma$ level) surface differential rotation.

\begin{figure}
   \centering
   \includegraphics[width=9cm]{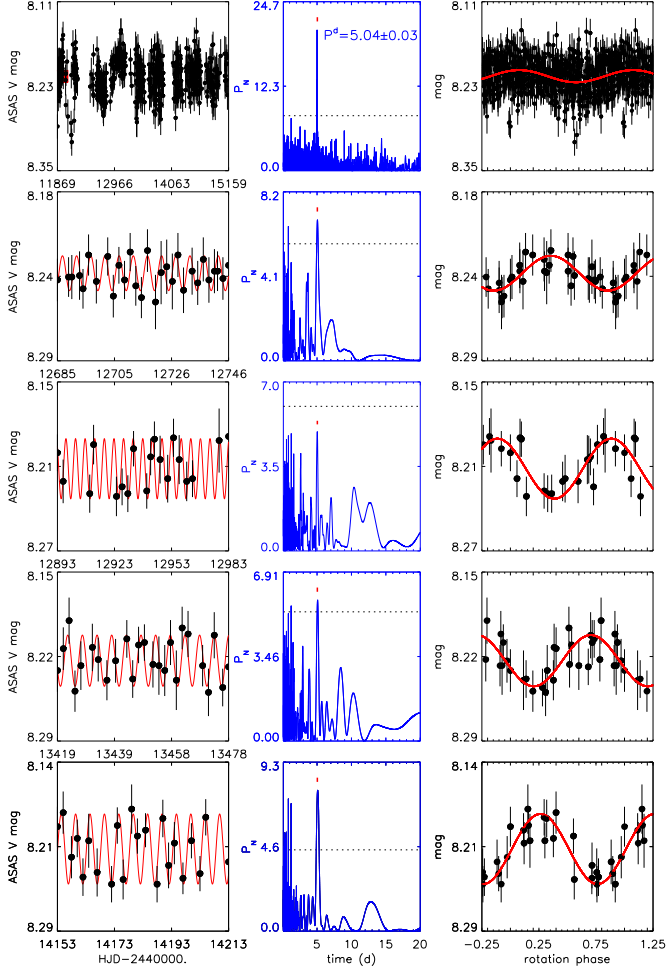}
   \caption{The complete ASAS magnitude time series  (top panel), together with a selection of time segments. 
In the left panels, we plot the time series magnitudes with an overplotted sinusoid (solid curve) of P=5.04d rotation period. 
In the middle panels, we plot the periodograms with indication of the 99\% confidence level (horizontal dashed line) 
and the power peak corresponding to the rotation period. Finally, in the right panels we plot the phased light curves
using the P=5.04d period and the first JD of the time segment as initial epoch. Here again the solid curve is the fitting 
sinusoid. }
              \label{f:prot}%
    \end{figure}

In Fig.~\ref{f:prot} we plot the complete ASAS magnitude time series  (top panel) together with a selection of time segments. 
In the left panels, we plot the timeseries magnitudes along with an overplotted sinusoid (solid curve) with a P=5.04d rotation period. 
In the middle panels, we plot the periodograms with indication of the 99\% confidence level (horizontal dashed line) 
and mark the power peak corresponding to the rotation period. Finally, in the right panels we plot the phased light curves
using the P=5.04d period and the first JD of the time segment as the initial epoch. Here again the solid curve is the fitting 
sinusoid. 
In the third panel from top, we notice that there may be time segments where the
light curve has a clear rotational modulation, although the power peak does not reach the 99\% confidence level.

\subsection{Long-term photometric variability}

\begin{figure}
   \centering
   \includegraphics[width=9cm]{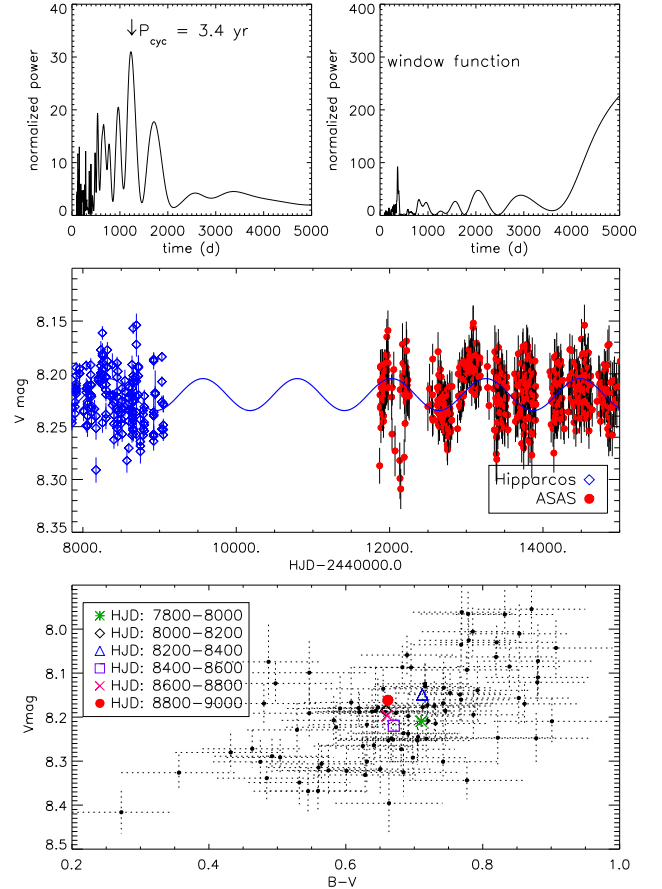}
   \caption{Top panels: the periodogram (left) and the window function
(right) obtained from the Lomb-Scargle analysis of the
ASAS plus Hipparcos time series. Middle panel: Hipparcos + ASAS V band magnitude timeseries 
with an overplotted fitted sinusoid of P=3.4yr cycle. Bottom panel: 
V versus B$-$V measurements from Tycho (individual measurements
as well as the binned values).}
   \label{f:cycle}%
    \end{figure}

The Lomb-Scargle periodogram analysis performed on the joined ASAS and Hipparcos datasets
also reveals a periodic long-term variability with a period of
P$_{\rm cycle}$=3.4$\pm$0.1 yr and a confidence level larger than 99\%.
The amplitude appears to change from cycle to cycle, as generally observed in cyclic spotted stars,
and reaches a maximum value of about A$_{\rm cycle}\simeq$0.08 mag.
In Fig.~\ref{f:cycle} we plot the Hipparcos + ASAS V band magnitude time series with an overplotted 
fitted sinusoid of P=3.4yr cycle.
Similar results are obtained when the periodogram analysis is carried out on only the ASAS dataset.
The Tycho magnitudes are excluded from this analysis, since their uncertainties are much greater 
than contemporary Hipparcos magnitudes.

The long-term variation probably represents a starspot cycle consisting of a quasi-periodic change of the star's spottedness 
similar to the 11-yr sunspot cycle.
It is very interesting to note that rotation period, cycle period, and cycle amplitude of HD\,61005 are all very similar
to the values observed in the Pleiades-age spotted star \object{DX Leo} that has P$_{\rm rot}$=5.42d, 
P$_{\rm cycle}$=3.1$\pm$0.05, and A$_{\rm cycle}\simeq$0.12 mag (Messina \& Guinan \cite{messina02}). 
Both stars are well placed  on   
the {\it active branch} observed by Saar \& Brandenburg (\cite{saar99})
for the distribution of the activity cycle frequency (normalized to the rotation frequency) versus inverse 
Rossby number for active stars.

Unfortunately, the B$-$V color from Tycho, which has quite a large  average uncertainty of $\sigma_{\rm B-V}$=0.15 mag, 
is the only one available to investigate the expected color variations of HD~61005 originating 
from magnetic activity.
We averaged the Tycho series of B$-$V measurements (after removing points with uncertainties above 0.20 mag) 
using time bins of 200 days in order to explore possible color variations. However, the inferred long-term 
B$-$V variations are found to be a factor 2.5 smaller than the average uncertainty. A Pearson linear correlation 
test shows such long-term B-V variations to be uncorrelated at a significante level $\alpha>50\%$ (cfr. Bevington \cite{bevington69}).
In the bottom panel of Fig.~\ref{f:cycle}, we plot the V versus B$-$V measurements from Tycho, along with the binned values. 
We notice that both individual measurements and average values show the tendency of the star to be bluer when  
fainter. Owing to the very poor quality of the available photometry, this evidence must be taken with great caution. 
If the observed trend is real, it may suggest that besides cool starspots, the observed photometric
variability of  HD\,61005 also arises from bright faculae (see, e.g., Messina \cite{messina08}).

\subsection{Photometric color}

Beside photometric colors in Hipparcos catalogs (B$-$V=0.734, V$-$I=0.78, from
Tycho and Star Mapper respectively), high-quality broad band photometry 
was performed by Menzies et al.~(\cite{menzies90}). It does not fully sample
stellar variability  (mean of 8 measurements); however, the V magnitude
(8.225$\pm$0.008) agrees very well with the long-term ASAS and Hipparcos
magnitudes (see above).
Therefore, in the following we adopt the B$-$V=0.751$\pm$0.003 and V$-$I=0.805$\pm$0.005
from  Menzies et al.~(\cite{menzies90}).
Near-infrared magnitudes are available from 2MASS (J=6.905$\pm$0.026; H=6.578$\pm$0.046
$K_S$=6.458$\pm$0.024). We obtain $V-K_{S}=1.762\pm0.025$.

\section{Determination of spectroscopic parameters}
\label{s:spec}


We exploited 12 spectra of HD 61005 taken in five nights between 
December 2003 and March 2006 using FEROS. These data were part of the RV survey for planets
around young stars by Setiawan et al.~(\cite{setiawan08}).
The FEROS spectra extend between 3600 \AA~ and 9200 \AA~ with a resolving 
power R = 48000 (Kaufer et al.~\cite{feros}). Three spectra were acquired 
using the object-sky set-up and the remaining using object-calibration mode
(see Table \ref{t:spec_feros}).
The data were reduced using a modified version of the FEROS-DRS pipeline 
(running under the ESO-MIDAS context FEROS), which yields 
the wavelength-calibrated, merged, normalized spectrum.
The reduction steps were the following:
bias subtraction and bad-column masking; definition of the echelle orders on
flat-field frames; subtraction of the background diffuse light; order extraction;
order by order flat fielding; determination of wavelength-dispersion solution 
by means of ThAr calibration-lamp exposures; 
order-by-order normalization, rebinning to a linear wavelength-scale 
with barycentric correction; and merging of the echelle orders. 

\subsection{Projected rotational velocity and radial velocity}

The radial-velocity (RV) and v$\sin{i}$ were determined by cross-correlating 
the object spectra with a low-v$\sin{i}$ template spectrum 
(i.e. a solar spectrum) obtained with the same instrument and reduced in 
the same way. 
The normalized target spectrum was preliminarily rebinned to a logarithmic 
wavelength scale 
($\Delta\ln\lambda \equiv \Delta v/c  = 5\times10^{-6}$) and split into six 
wavelength-ranges that are free of emission lines and of telluric absorptions
(cf. Esposito et al. \cite{esposito07}). 
A Gaussian was fitted to the peak of the cross-correlation function (CCF), 
computed in each of the six distinct spectral ranges. The resulting RV values were 
averaged with weights (Schisano et al. \cite{schisano09}). 
The projected rotational velocity v\,$\sin{i}$ was derived from the FWHM of 
the cross-correlation peak in each of the six ranges above.
The relation FWHM-vsini was obtained by convolving the template spectrum
with rotational profiles of vsini from 1 to 100 km/s.
Our measurements are listed in Table \ref{t:spec_feros}.

In Table \ref{t:spec_lit} we collect the measurements of $v \sin i$ available for HD\,61005.
All the determinations except the one by White et al.~(\cite{white07}), which is based on data with lower spectral resolution,
show fairly good agreement. In the following analysis we adopt our own determination
($v \sin i=8.2\pm0.5$~km/s).

\begin{table*}[h]
\caption{Measurements of FEROS spectra. OS= Object+Sky OC=Object=Calibration (simultaneous thorium). } 
\label{t:spec_feros}
\begin{center}       
\begin{tabular}{ccccccl} 
\hline
JD         & RV  &  vsini   &  Teff & EW Li &  S index &  Remarks \\
-2450000   & km/s & km/s    &  K    & m\AA  &          &          \\
\hline
2982.81810 & $+22.5\pm0.1$ & $7.8\pm0.8$ & $5521.11\pm22.42$  &    163. &  0.492 & OS \\
2982.82576 & $+22.5\pm0.1$ & $7.8\pm0.8$ & $5543.06\pm33.52$  &    170. &  0.501 & OC \\
3447.62607 & $+22.5\pm0.1$ & $8.4\pm0.8$ & $5496.60\pm22.20$  &    172. &  0.518 & OS \\
3447.63732 & $+22.5\pm0.1$ & $8.4\pm0.8$ & $5488.43\pm24.10$  &    173. &  0.529 & OC \\
3807.55698 & $+22.5\pm0.1$ & $8.4\pm0.9$ & $5487.95\pm26.98$  &    173. &  0.520 & OS \\
3807.61385 & $+22.5\pm0.1$ & $8.5\pm0.9$ & $5497.87\pm28.86$  &    172. &  0.508 & OC \\
3807.69345 & $+22.5\pm0.1$ & $8.4\pm0.9$ & $5485.21\pm28.06$  &    171. &  0.506 & OC \\
3810.50830 & $+22.5\pm0.1$ & $8.2\pm0.8$ & $5511.77\pm23.01$  &	   171. &  0.512 & OC \\ 
3810.58558 & $+22.5\pm0.1$ & $8.2\pm0.9$ & $5495.18\pm23.51$  &	   172. &  0.517 & OC \\
3810.70741 & $+22.5\pm0.1$ & $8.3\pm0.8$ & $5477.39\pm25.16$  &	   173. &  0.508 & OC \\
3815.56421 & $+22.5\pm0.1$ & $8.2\pm0.8$ & $5481.98\pm24.40$  &    173. &  0.507 & OC \\
3815.64599 & $+22.5\pm0.1$ & $8.2\pm0.8$ & $5495.80\pm22.50$  &    172. &  0.506 & OC \\
\hline
mean       &   22.5        &     8.23    &  5499              &  171.3  &  0.510 & \\ 
rms        &    0.0        &     0.27    &    19              &    2.8  &  0.010 & \\
\hline
night mean &   22.5        &     8.21    &  5500              &  171.1  &  0.510 & \\
rms        &    0.0        &     0.25    &    18              &    2.6  &  0.010 & \\
\hline
\end{tabular}
\end{center}
\end{table*}

\begin{table*}[h]
\caption{Summary of measurements of spectroscopic parameters of HD~61005 available in the literature} 
\label{t:spec_lit}
\begin{center}       
\begin{tabular}{cccclcl} 
\hline
RV  &  vsini   & EW Li & S index &   Instrument & Resolution  & Reference \\
\hline
               &                 &             &  0.492           &            &            & Henry et al.~\cite{henry96} \\
22.0           &  8              & 176         &                  & FEROS      &  48000     & Wichmann et al.~\cite{Wichmann03} \\
$22.3\pm0.2$   &  9              &             &                  & CORAVEL    &            & Nordstrom et al.~\cite{nordstrom04} \\
23.0           &  8              &             &                  & AAT        &  51000     & Waite et al.~\cite{waite05} \\
               &                 &             &  0.459 	  & CTIO 1.5m  &   1100     & Gray et al.~\cite{gray06} \\
$21.98\pm0.85$ &  14.9           & 176         &  $0.444\pm0.081$ & Palomar    &  16000     & White et al.~\cite{white07} \\
$22.5\pm0.1$   &                 &             &                  & FEROS      &  48000     & Setiawan et al.~\cite{setiawan08} \\
               &  8.2            &             &  0.471           & FEROS      &  48000     & Schroeder et al.~\cite{schroeder09} \\
               &  $9.9\pm0.9$    & $169\pm3$   &                  & FEROS      &  48000     & Weise et al.~\cite{weise10} \\
\hline 
$22.5\pm0.1$   &  $8.2\pm0.5$  & $171\pm3 $ &  0.516           & FEROS      &  48000     & this work \\
\hline 
$22.5\pm0.1$   &  $8.2\pm0.5$  & $171\pm3 $ &  0.502           &            &            & adopted \\
\hline
\end{tabular}
\end{center}
\end{table*}

\subsection{T$_{\rm eff}$ determination}
\label{s:teff}

\begin{figure}
   \centering
   \includegraphics[width=6cm,angle=-90]{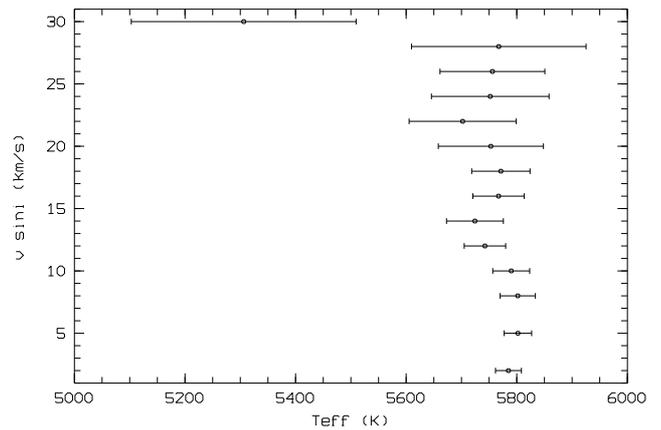}
   \caption{Effective temperature derived with ARES on Sun spectra broadened artificially 
            to various projected rotational velocities. 
	     Results are robust for $v \sin i$ up to 18 km/s.}
              \label{f:teffsun}%
    \end{figure}

The determination of the effective temperature was obtained through the method
of equivalent widths (EW) ratios of spectral absorption lines, using the calibration for FGK 
dwarf stars by Sousa et al.~(\cite{sousa10}). The EWs were measured 
using the ARES\footnote{http://www.astro.up.pt/~sousasag/ares/}
automatic code (Sousa et al.~\cite{sousa07}).
In order to test the performances of ARES on spectra with different
rotational broadening, we used a FEROS spectrum of the Sun broadened
artificially with rotation profiles of increasing $v \sin i$.
The resulting values of effective temperature remain consistent (within
~50K) with the accepted value for the Sun up to a rotational velocity of about
18 km/s (Fig.~\ref{f:teffsun}).
For higher rotation velocity, the number of measured
line-ratios drops drastically with correspondingly larger errors due to increased line
blending.
The temperature determinations become unreliable for $v \sin i$ approaching 30~km/s.

The value of T$_{\rm eff}$ determined by ARES (5550~K) is consistent with those
derived from excitation equilibrium as part of the abundance analysis 
(5500$\pm$60~K; Sect.~\ref{s:abu}), from B$-$V, V$-$I and V$-$K colors using the
calibration by Casagrande et al.~\cite{casagrande10} (5480~K), 
the relationship between W(Na\,I\,D) vs. T$_{\rm eff}$ by 
Tripicchio et al.~(\cite{Tripicchio97}) (5550~K), and the previous
determination by Hines et al.~(\cite{hines07}) (5456~K from SED fitting
using Kurucz models).

\subsection{Abundance analysis}
\label{s:abu}

There was no previous abundance determination based on high resolution spectra.
Gray et al.~(\cite{gray06}) derived [M/H]$=-0.06$ from modeling of their low resolution 
spectrum. The lower value derived by Nordstrom et al.~(\cite{nordstrom04}) from
Str\"omgren photometry ([Fe/H]$=-0.19$) is biased by the effect of stellar activity
(see Favata et al.~\cite{favata97}).

We performed elemental abundance analysis for HD~61005, measuring
abundances for several iron-peak and $\alpha$-elements, namely Fe, Ni, Na, Si, Ca, Ti,
and the $s$-process element Ba.
The chemical analysis was carried out following the same approach and methodology
as in D'Orazi \& Randich (\cite{dorazi09}, hereafter DR09). 
The Ba content was instead derived as in D'Orazi et al.~(\cite{dorazi09a}), taking 
isotopic splitting and hyperfine structure into account for both Ba~{\sc ii} features at
5853 \AA~and 6496 \AA~(see D'Orazi et al.~\cite{dorazi09a} for further details on this issue).

Our procedure can be briefly summarized as follows.
We employed the line list provided in DR09,
deriving LTE force-fitting abundances with the driver {\it abfind} in MOOG 
(Sneden \cite{sneden73}, 2002 version)
and the Kurucz (\cite{kurucz93}) set of model atmospheres.
The EWs were measured with the IRAF task {\sc SPLOT}, performing a Gaussian fit to
line profile. All the features with EW values higher than $\sim$150 m\AA~were
discarded to avoid spectral lines close to the saturation regime of the curve
of growth.

Stellar atmospheric parameters were spectroscopically optimized:
effective temperature (T$_{\rm eff}$) and microturbulence ($\xi$)
values were obtained zeroing the slope of $\log$~n(Fe~{\sc i}) with the
excitation potential ($\chi$) and with line strengths, respectively.
Concerning gravity, we imposed the ionization equilibrium condition,
i.e., log~n(Fe{\sc i})$-$log~n(Fe{\sc ii})$\leq$0.05 dex,
to derive the log~g value.

We obtained 
T$_{\rm eff}$=5500$\pm$60K, $\xi$=1.00$\pm$0.25 km/s, and
$\log$~g=4.5$\pm$0.2 as final adopted parameters. We refer the reader to DR09 for a detailed discussion on
error estimates; here we just mention that internal errors, due to EW
measurements and atmospheric parameters, are not larger than 0.08 dex.
As to systematic (external) errors, our analysis (with the same method, code,
line list) of the Hyades star VB~187 indicates
that no major uncertainties affect our abundance estimates (see DR09).
In Table~\ref{t:abu} [Fe/H] and [X/Fe] ratios are reported for HD~61005 along with
internal uncertainties due to EW values.
The [X/H] ratios were computed by assuming the ones derived by
DR09  as solar abundances.  
As a comparison, average abundances from seven stars in the open cluster \object{IC 2391} 
are given and
the standard deviation from the
mean (rms) is provided in correspondence of each [X/Fe] mean ratio. 

\begin{table*}
\begin{center}
\caption{Metallicity and [X/Fe] ratios for HD~61005, where errors represent
uncertainties due to EWs.}
\label{t:abu}
\setlength{\tabcolsep}{1.45mm}
\begin{tabular}{lccccccc}
\hline
Name & [Fe/H] & [Na/Fe] & [Si/Fe] & [Ca/Fe] & [Ti/Fe] & [Ni/Fe] & [Ba/Fe] \\
     &        &         &         &         &         &         &        \\
\hline
HD~61005 & ~~0.01$\pm$0.04 & $-$0.03$\pm$0.05 & ~~0.00$\pm$0.04 & ~~0.00$\pm$0.04 &
~~0.02$\pm$0.06 & ~~0.00$\pm$0.06 & ~~0.63$\pm$0.06 \\
IC~2391$^{a}$  & $-$0.01$\pm$0.02 & $-$0.04$\pm$0.05 & ~~0.01$\pm$0.02 & ~~0.02$\pm$0.01 &
~~0.00$\pm$0.02 & ~~0.00$\pm$0.02 & ~~0.68$\pm$0.07\\

\hline
\end{tabular}
\end{center}
$^{a}$ Average abundances (with standard deviations from the mean)
for IC~2391 by D'Orazi \& Randich \cite{dorazi09} and D'Orazi et al.~\cite{dorazi09a}, for comparison
\end{table*}

\subsection{Lithium}

The EW of the Li 6708~\AA~~resonance line was determined by Gaussian fitting with the  ARES code.
(see Table~\ref{t:spec_feros}).
Our result (EW=171~m\AA) agrees closely with the determinations available in the 
literature (Table \ref{t:spec_lit}).

The Li~{\sc i} feature at 6707.78 \AA~is an unresolved doublet so we conducted a spectral 
synthesis analysis to derive a reliable Li abundance.
Using the driver {\it synth} in MOOG and Kurucz model atmospheres 
(as previously done for our abundance analysis, see Sect.~\ref{s:abu}),
we computed synthetic spectra in a wavelength window of $\sim 15$ \AA, from 6700 \AA~to 6715 \AA,
and employed the line list given in D'Orazi et al.~(\cite{dorazi09b}).
As a first step, we optimized the line list by changing the oscillator strengths, 
i.e. $\log$~$gf$ values, when necessary, to obtain the best agreement between
our FEROS solar spectrum and the standard iron abundance by Anders \& Grevesse (\cite{anders89}).
To perform the comparison between
the observed spectrum of HD~61005 and the synthetic ones,
we convolved the latter with both a Gaussian profile at our resolution of R$\sim$48000 (FWHM=0.14 \AA)
and a rotational profile, taking the limb-darkening coefficient into account.

Concerning atmospheric parameters, for consistency we adopted the ones derived from our EW 
abundance analysis, namely T$_{\rm eff}$=5500K,
log g=4.5, $\xi$=1.0 kms$^{-1}$, and [Fe/H]=0 dex.
As a by-product, the spectral synthesis allowed us to infer the projected
rotational velocity and we obtained v$\sin$i=9.0$\pm$0.5 kms$^{-1}$, in perfect agreement 
with other available estimates (see Table \ref{t:spec_lit}).

\begin{figure}
   \centering
   \includegraphics[width=8.5cm,height=7cm]{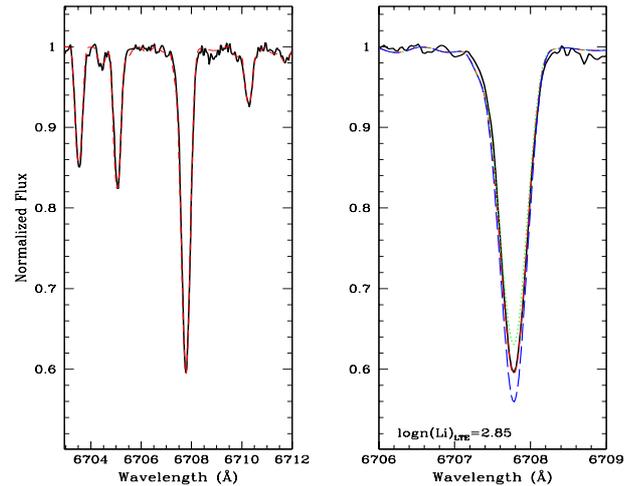}
   \caption{Left panel: Spectral synthesis of the region around the Li
             6708~\AA~~doublet.
             Right panel: details of the Li doublet. Overplotted is
              our best-fit Li abundance (log~n(Li)=2.85; red short-dashed curve)
            while the dotted and long-dashed curves represent log~n(Li)=2.75 
            and log~n(Li)=2.95, respectively.}
              \label{f:lisynt}%
    \end{figure}

As shown in Fig.~\ref{f:lisynt}, all the observed features are reproduced very well
by the synthetic profiles, which simultaneously fit the Fe~{\sc i} lines (6703, 6704, and 6710 \AA) 
and the Li~{\sc i} doublet. The best agreement between observed and synthetic spectra
is achieved for a Li abundance of log~n(Li)=2.85 (righthand panel 
of Fig.~\ref{f:lisynt}).

Uncertainties related to the best-fit determination and to the adopted T$_{\rm eff}$
 are 0.05 dex and 0.07 dex, respectively (note that Li abundance is only scarcely affected by other stellar parameters, 
i.e., log g and $\xi$);
to estimate the total error, we quadratically summed both these contributions, given 
their independence, and we found 0.09 dex. In passing, we also recall that uncertainties 
due to non LTE (NLTE) and/or three-dimensional model atmospheres
can be neglected, since they cancel each other (see Asplund \& Lind \cite{asplund10} for details).

\subsection{Chromospheric emission}
\label{s:hk}

Chromospheric emission in CaII H\&K lines of HD~61005 was determined
by measuring the $S$ index on FEROS spectra as in Desidera et al.~(\cite{desidera06}).
The $S$ index was also measured on 73 FEROS spectra of 41 stars from the lists of 
Baliunas et al.~(\cite{baliunas95}) and Wright et al.~(\cite{wright04}), reduced in the same
way as those of HD\,61005, to
calibrate our instrumental $S$ index onto the standard M. Wilson scale.
The dispersion of the residuals of the calibration is 0.026, most likely dominated by
the intrinsic variability of chromospheric emission.
The variability of the $S$ index measured on 15 spectra of the chromospherically 
quiet star \object{$\tau$ Ceti} is
0.0034 (2.0\%). Full details of the calibration will be provided in a forthcoming paper.
There are other four measurements of H\&K emission of HD 61005 in the literature listed in
Table \ref{t:spec_lit}. The average of our nightly averages and the literature determinations
(excluding those by Gray et al.~\cite{gray06} and White et al.~\cite{white07} because of lower accuracy)
yields a mean $S$ index of 0.502 (rms 0.017), which corresponds to $\log R'_{HK}=-4.310$.
The expected rotational period for a star with such activity level and color
using the calibration by Mamajek \& Hillenbrand~(\cite{mamajek08}) is 5.01d, very close to the
observed one (5.04d). 

\section{System parameters}
\label{s:param}

\subsection{Photometric and trigonometric distances}

\begin{figure}
   \centering
   \includegraphics[width=8.5cm]{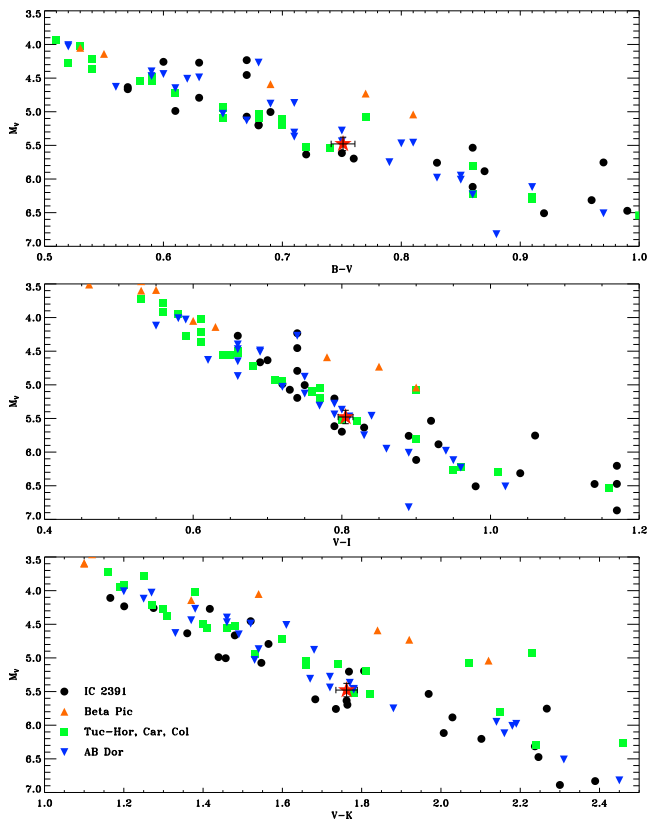}
   \caption{$M_{V}$ vs B-V, V-I  and V-K color magnitude diagrams for HD 61005 (red star),
    IC 2391 open cluster (black filled circles; data summarized in Messina et al.~\cite{messina11}), 
    $\beta$ Pic moving group (orange filled triangles, 10 Myr),
    Tucana, Carina and Columba associations (green filled squares, 30 Myr) and AB Dor moving group 
    (blue filled, upside-down triangles, 70 Myr). For these  associations, only the stars with trigonometric
   parallaxes among the members selected by Torres et al.~(\cite{torres08}) were considered.
     The position of HD 61005 is compatible with a broad range 
     of ages, older than about 20 Myr.}
              \label{f:cmd}%
    \end{figure}

The trigonometric parallax in Van Leuveen (\cite{vanlee07}) is $28.29\pm0.85$ mas.
The position in the color-magnitude diagram agrees with those of the members of other
groups in the age range 30-100 Myr, while ages under 20 Myr (see Fig.~\ref{f:cmd})
are excluded.
Estimated photometric distances (from B-V, V-I and V-K colors) using the sequences of 
\object{AB Dor} and \object{Tucana} associations are 37 and 39 pc, respectively, 
a not significantly different from the trigonometric one 
(35.3$\pm$1.1 pc, Van Leuveen \cite{vanlee07}), which we adopt in the following.

\subsection{T$_{\rm eff}$, luminosity, and radius}

Using the absolute visual magnitude $M_{V}=5.478$ and the bolometric correction $-0.162$ from
Flower (\cite{flower96}), we derive a luminosity $L=0.583~L_{\odot}$.
Coupled with the adopted effective temperature of T$_{\rm eff}$=5500~K (Sect.~\ref{s:teff}), 
this gives a stellar radius of $R=0.840 \pm 0.038 ~R_{\odot}$ (compatible within
errors with the determination by Watson et al.~\cite{watson10}; $R=0.829\pm0.048$).

\subsection{System inclination}

The measured rotational period of 5.04d coupled with the stellar radius
of $0.840\pm0.038~R_{\odot}$ yields a rotational velocity of $8.43\pm0.38$ km/s.
Coupled with our adopted projected rotational velocity of $v sini=8.2 \pm 0.5$ km/s,
this gives $ \sin i=0.97\pm0.09$ and $i=77^{\circ}$$^{+13}_{-15}$.
The ring inclination is $84.3^{\circ}\pm1$ (Buenzli et al.~\cite{buenzli10}), with a corresponding
$\sin i$ of 0.995.

We conclude that both the star and the disk are viewed close to edge-on and that
there are no indications of any misalignment.
A similar conclusion was reached by Watson et al.~(\cite{watson10}). However, their analysis is
based on the expected rotation period from H\&K emission and the Noyes et al.~(\cite{noyes84}) relation
rather than the measured value, which was derived for the first time in our study.

\subsection{X-ray emission}
\label{s:xray}

The star is present neither in the ROSAT all-sky survey (RASS)
catalog nor in other X-ray catalogs. We note that there are
no sources in the official RASS catalogs within 1.2 deg from
HD 61005. This is unexpected considering the young age
inferred from other indicators and the distance from the Sun.
However, in the paper by Wichmann et al.~(\cite{Wichmann03}),
an X-ray luminosity $\log L_{X}=29.2$ for the star is quoted,
which was based on a preliminary RASS catalog, which is older than the
official one. The source is not in the official catalog most
likely because it may have been discarded as spurious in the
final RASS data screening (Wichmann 2010, private communication).
Therefore, the original RASS data have been kindly revisited
by J\"urgen Schmitt. The source is clearly recognizable 
by eye.
The EXSAS\footnote{EXSAS stands for Extended Scientific Analysis
System to evaluate RASS data; see Zimmermann et al.~(\cite{Zimmermann94})
for details} source detection algorithm gives a broad-band count rate of
0.176$\pm$0.034\,cts/sec (32.21$\pm$6.27 counts in 272.4 sec exposure
time, and vignetting correction factor of 1.493) and a hardness ratio
practically equal to zero. Adopting the calibration by Hunsch et al.
(1999)
and the trigonometric distance, an X-ray luminosity
$\log L_{X}=29.3 \pm 0.2$ is derived.

\subsection{Radial velocity, binarity, and planetary companions}

There are no indications of any RV variations with amplitudes larger than 1 km/s
from the various literature data or from our own RV measurements based on FEROS data. 
We therefore exclude the star being is a spectroscopic binary.
The RV curve by Setiawan et al.~(\cite{setiawan08}) shows a significant scatter
with peak-to-peak amplitude of about 150 m/s and a correlation with line profile.
Therefore, the RV variability appears dominated by the line profile alterations caused by stellar
activity, and the limits on close-in planets are rather coarse, excluding only massive
planets in close orbits.  
The deep imaging also rules out either stellar, brown dwarf, or even massive 
planetary companions at a projected 
separation larger than a few tens of AU (Buenzli et al.~\cite{buenzli10}).
The similarity between short-term Hipparcos and historical  proper motions also
point to a single star.
Overall, there are no indications of any massive companions at either short or wide
separations. The eccentricity of the ring suggests a giant planet within a few tens of AU
(Buenzli et al.~\cite{buenzli10}).

\subsection{Kinematic parameters}
\label{s:kin}

\begin{figure}
   \centering
   \includegraphics[width=6.0cm,angle=-90]{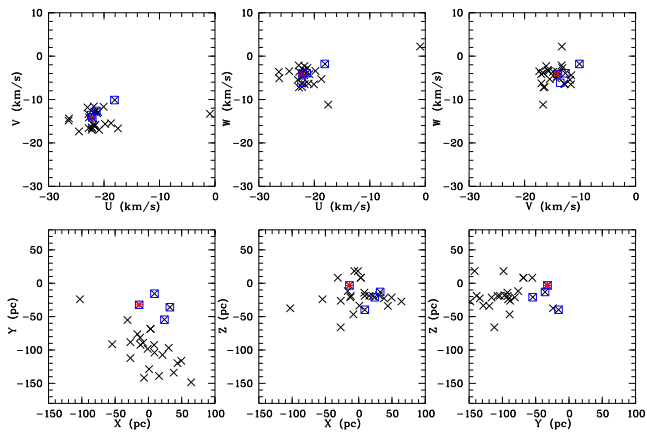}
   \caption{Upper panels: U,V, and W space velocities for HD 61005 (red asterisk)
  and Argus members from the list of Torres et al.~\cite{torres08}. Square symbols 
  refer to objects with trigonometric parallaxes.
  Lower panels: space positions X, Y, Z. Symbols as in the upper panels.}
              \label{f:uvw}%
    \end{figure}

The space velocities of HD~61005 are $U=-22.2\pm0.6$~km/s, $V=-14.3\pm0.3$~km/s, and
$W=-4.1\pm0.2$~km/s.
These are very similar to those of the \object{Argus association}
($-22.0\pm0.3$ $-14.4\pm1.3$ $-5.0\pm1.3$) as given in Torres et al.~(\cite{torres08}).
The space position is $X=-13.9\pm0.4$ pc, $Y=-32.3\pm1.0$ pc, and $Z=-3.3\pm0.1$ pc, 
which places the star at the outskirts of the known members of the Argus association 
(Fig.\ref{f:uvw}).

The existence of the  Argus association was first proposed by  
Torres et al.~(\cite{torres03}) from SACY data, thanks to its peculiar $U$ velocity.
The young open cluster IC~2391 was shown to share the motion and the age
of the proposed association.
The occurrence of groups of young field stars associated with IC 2391 had been
formerly proposed by Eggen (\cite{eggen91}) and Makarov \& Urban (\cite{makarov00}); however,
only a minor fraction of their proposed members is part of the Argus association
according to Torres et al.~(\cite{torres08}).
 
Torres et al.~(\cite{torres08}) give an updated list of members of
this association separated as field stars and IC\,2391 members
(29 and 35 stars, respectively, see also the review by Pettersson~\cite{petter08}).
The estimated age of the association is 40 Myr, in the middle of the proposed ages 
for IC 2391 (30$\pm$5\,Myr Stauffer et al.~\cite{stauffer97} using isochrone fitting,
50$\pm$5\,Myr Barrado y Navascues et al.~\cite{barrado04} using the Li depletion 
boundary, 30-50\,Myr Platais et al.~\cite{platais07}).  

The mean distance of the members in Torres et al.~(\cite{torres08}) is 106 pc, much farther
than the 35 pc of HD~61005, but a few individual members have
comparable distance. The trigonometric distance to IC 2391 is $144.9\pm2.5$ pc (Van Leeuwen \cite{vanlee09}), 
and it is expected that several field
members at distances  comparable to or larger than IC 2391 wait to be identified 
(assuming a roughly spherical spatial distribution).

HD\,61005 is not included in the member list by Torres et al.~(\cite{torres08}).
This  can be easily understood, because the SACY sample (Torres et al.~\cite{torres06}) is based 
on a cross-identification of the ROSAT All Sky Bright Source Catalog (Voges et al.~\cite{rosatbright}), 
of which HD\,61005 is not included (see Sect.~\ref{s:xray}).

The young age clearly resulting from all the indicators described above and 
the very similar space velocity call for a detailed evaluation of the possible
membership of HD\,61005 to the Argus association.
This is the goal of the next section.

\section{The age of HD\,61005 and the membership to Argus association}
\label{s:age}

In the previous section we measured or collected from the literature the determination
of several parameters that are known to depend on stellar age.
We discuss here whether they are compatible with the Argus membership suggested by
kinematic properties or if they instead indicate a different age.
Application of existing age calibrations is summarized in Table \ref{t:age}.
The calibration of lithium EW into age was built using literature measurements of
nearby young associations and Pleiades and Hyades open clusters and will be presented
elsewhere.

\begin{table}[h]
\caption{Summary of age determination for HD~61005} 
\label{t:age}
\begin{center}       
\begin{tabular}{lcl} 
\hline
Indicator &  Age           & Calibration  \\
\hline
Li        &   113          & see text   \\
H\&K      &   148          & Mamajek \& Hillenbrand (\cite{mamajek08}) \\
H\&K      &    72          & Donahue (\cite{donahue93}); Henry et al~(\cite{henry96}) \\
Xray      &   108          & Mamajek \& Hillenbrand (\cite{mamajek08})  \\
P         &   186          & Mamajek \& Hillenbrand (\cite{mamajek08}) \\
P         &   125          & Barnes (\cite{barnes07}) \\
\hline 
\end{tabular}
\end{center}
\end{table} 

The ages in Table \ref{t:age} are in general older than the accepted age of the Argus 
association\footnote{Weise et al.~(\cite{weise10}) estimated an age of 30 Myr for HD~61005 
from Li EW. This is partially due to the younger ages assumed for the clusters adopted
as calibrators.}. At  first sight, this argues against membership.
However it is known that most (possibly all) of the HD 61005 age indicators have some correlation 
with stellar rotation, and show a wide dispersion in coeval (especially young) 
stellar groups and clusters. A one-to-one correspondence between color and rotation period is 
definitively reached by an age of about 500-600 Myr (see, e.g., Collier Cameron et al. \cite{cameron09}).
The dispersion in rotation rate for coeval systems might be linked to the disk lifetime
(disk-locking scenario). The active debris disk of HD 61005 might be a relic of a long-lived
primordial disk. This might have prevented the system reaching fast rotation rates, increasing at the
same time the lithium depletion (Bouvier \cite{bouvier08}).
A more detailed discussion is therefore needed to see if there are any known Argus members that are
slow rotators and present similar discrepancies in the age indicators
to HD~61005.
Furthermore, we also consider here additional membership criteria such as chemical composition.

\subsection{Rotation period}

The rotation period of HD\,61005 was determined in Sect.~\ref{s:prot}, while
those of Argus and IC2391 members are taken from Messina et al.~(\cite{messina11}).
They classified the detection of rotation period as either confirmed, likely, or
uncertain depending on the number of significant detections of the periodicity among the 
segments by which the photometric time series were divided.
We use this classification in the following discussion.
The four stars (CD-582194 and CD-621197 in Argus, PMM351 and PMM 2182 in IC2391) 
for which the rotation period is inconsistent with the spectroscopic 
$v \sin i$ are not used in the analysis. 

\begin{figure}
   \centering
   \includegraphics[width=9cm]{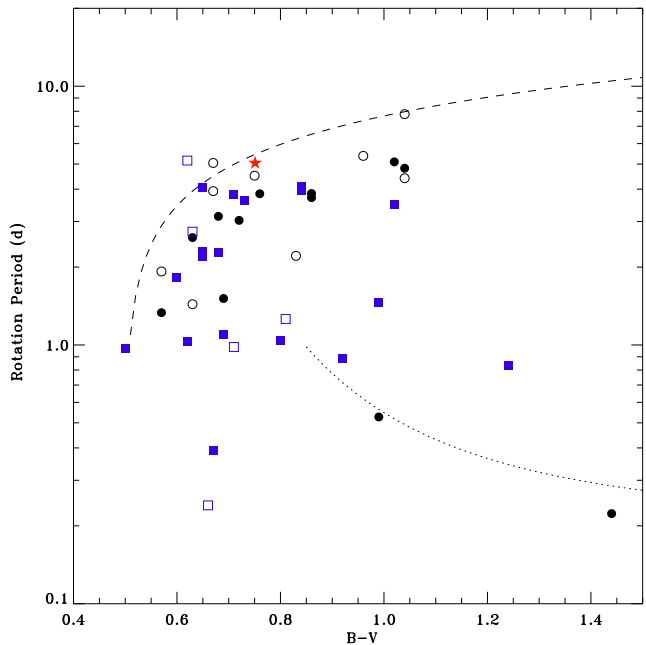}
  \caption{Upper panel: distribution of rotation periods of Argus association 
           (blue squares) and IC 2391 (black circles) 
            members versus B$-$V color. Confirmed and likely periods are plotted with filled symbols, 
            uncertain periods with open symbols. Data are taken from Messina et al.~(\cite{messina11}). 
            HD~61005 is represented by the (red) star symbol. 
            The dashed and dotted lines represent 
            the upper bounds of the period distribution of 
            Pleiades members (from Barnes \cite{barnes03}). } 
              \label{f:hd61005argusrot}%
    \end{figure}

The rotation period of HD\,61005 compared to that of members of Argus (field) and IC2391
is shown in Fig.~\ref{f:hd61005argusrot}.
The distribution of rotation periods of IC 2391 stars appears different from that of Argus field
stars, with more slow rotators in the cluster.
This might be from the different selection criteria of the IC 2391 cluster and the Argus association field members.
Argus field members are identified in the sample of the SACY survey (Torres et al.~\cite{torres06})
and, by definition, have X-ray counterparts in the ROSAT Bright Source Catalog
(Voges et al.~\cite{rosatbright}).
Conversely, the members of IC 2391 are selected from a variety of sources, including deep
X-ray imaging of the central regions of the cluster (Patten \& Simon \cite{patten96}; 
Simon \& Patten \cite{simon98}; 
Marino et al.~\cite{marino05}) or a combination of spectroscopic, astrometric and photometric
criteria, without X-ray preselection (Platais et al.~\cite{platais07}).
Indeed, several members identified by Platais et al.~(\cite{platais07}) only have X-ray counterparts 
in the  ROSAT Faint Source Catalog (Voges et al.~\cite{rosatfaint}),
and a few targets have no X-ray counterparts at all (see App.~\ref{s:xray_ic2391}).
Since fast rotators have brighter coronal luminosities and the mean distance of Argus association
members is 106 pc, we expect that the census of association members is biased toward the most
active and fast-rotating stars.  
We therefore consider the IC~2391 cluster members as a better comparison sample than 
Argus field members
to evaluate the possible link of HD\,61005 with this association. 

The position of HD~61005 in the color-period plot of Fig.~\ref{f:hd61005argusrot} is 
on the margin of the IC~2391 period distribution, 
close to a few members (PMM665, PMM1373, PMM4362, and PMM6974) whose rotation period still need
confirmation. HD~61005 appears to be along the so-called ``interface sequence''  
that represents the upper bound of the period 
distribution. Here this sequence is represented by  the gyro-isochrone  
from  Barnes (\cite{barnes03}) and computed for the Pleiades age of 120 Myr.

We further note that the photometric amplitude is within the distribution  of
Argus and IC 2391 members with similar rotational period 
(see right panel of Fig.~6 in Messina et al.~\cite{messina11}).

\subsection{Lithium}

\begin{figure}
   \centering
   \includegraphics[width=9cm]{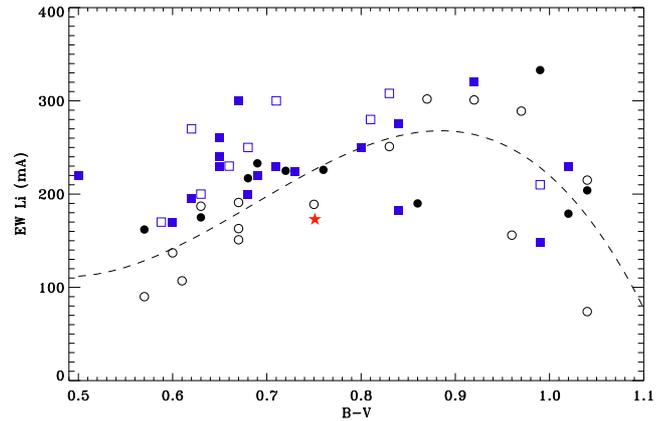}
   \caption{Comparison of Li EW of HD~61005 (red filled star) with that of Argus
   association field stars (blue squares) and  IC 2391 open cluster stars (black circles)
  Filled symbols refer to stars with confirmed or likely rotational period in
  Messina et al.~(\cite{messina11}), and open symbols to stars with uncertain
  or undetermined rotational period in Messina et al.~(\cite{messina11}).
  The dashed curve is the cubic fit to the locus of IC~2391 members in the range $B-V=0.50-1.10$}
              \label{f:ewli}%
\end{figure}

The comparison of Li EW of HD\,61005 with that of Argus association members 
(Torres et al.~\cite{torres06}) shows that
HD\,61005 has an EW about 50 m\AA~lower than that of members of similar color 
(Fig.~\ref{f:ewli}).
When one instead considers members of the open cluster  IC 2391
(from Platais et al.~\cite{platais07}; Randich et al.~\cite{randich01}; Stauffer et al.~\cite{stauffer97}), 
HD\,61005 falls on the edges of the distribution.
The Li EW also fits within the distribution
of the Pleiades observations (age about 120 Myr). Application
of an age-EW Li-color relation based on data of nearby young associations and Pleiades
and Hyades open clusters yields 113 Myr.

It appears that, in spite of the association between IC 2391 and the Argus field stars,
IC 2391 members have on average a slightly smaller lithium content.
At first sight this can be considered a signature of a small age difference between cluster and
field members. However, considering the possible dependence of lithium abundance on rotational 
period or rotational velocity found by Soderblom et al.~(\cite{soderblom93}) 
for the Pleiades and by Da Silva et al.~(\cite{dasilva09}) for members of young associations, 
this may just be a result of the selection effects described above.

\begin{figure}
   \centering
   \includegraphics[width=9cm]{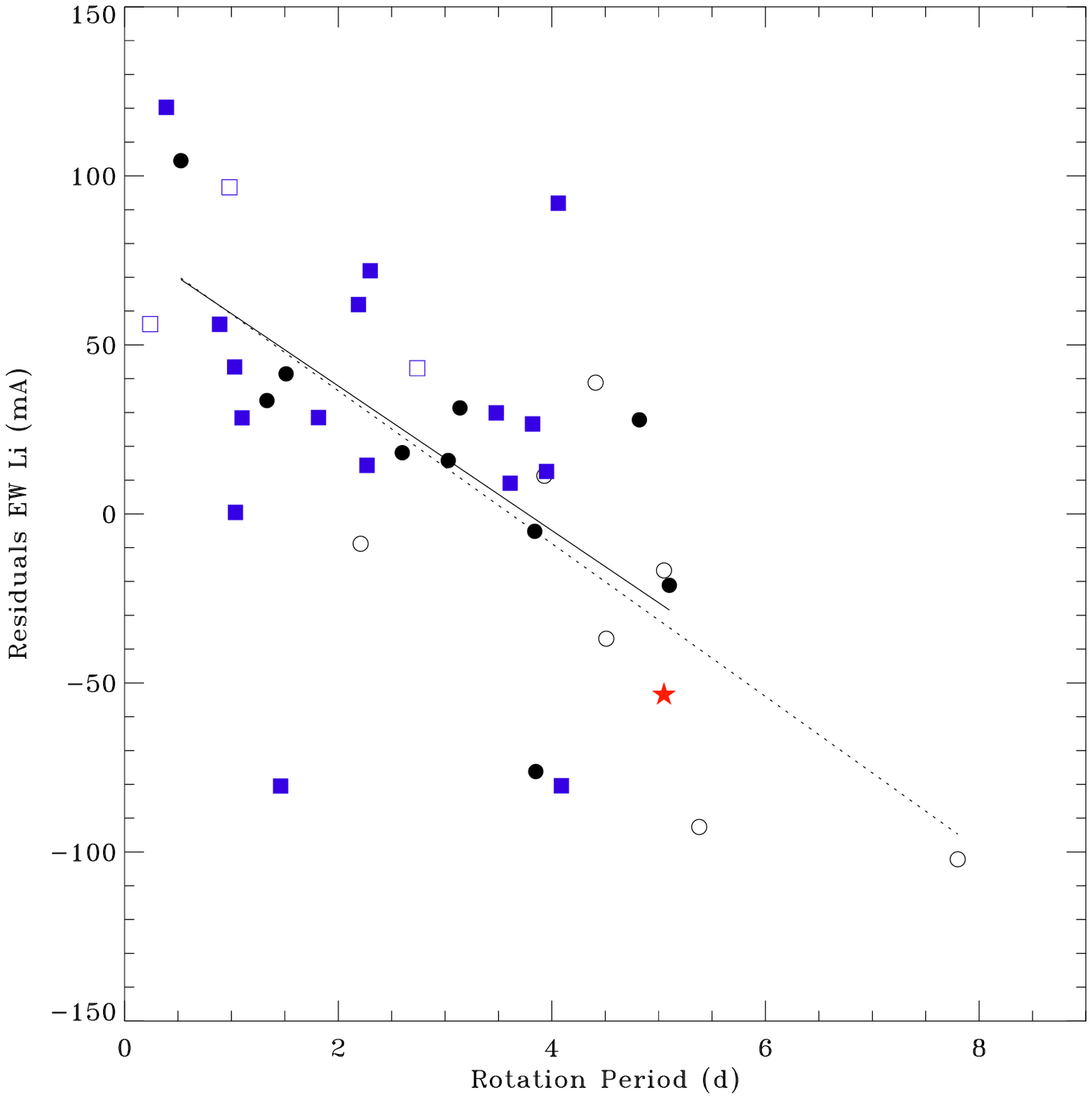}
      \includegraphics[width=9cm]{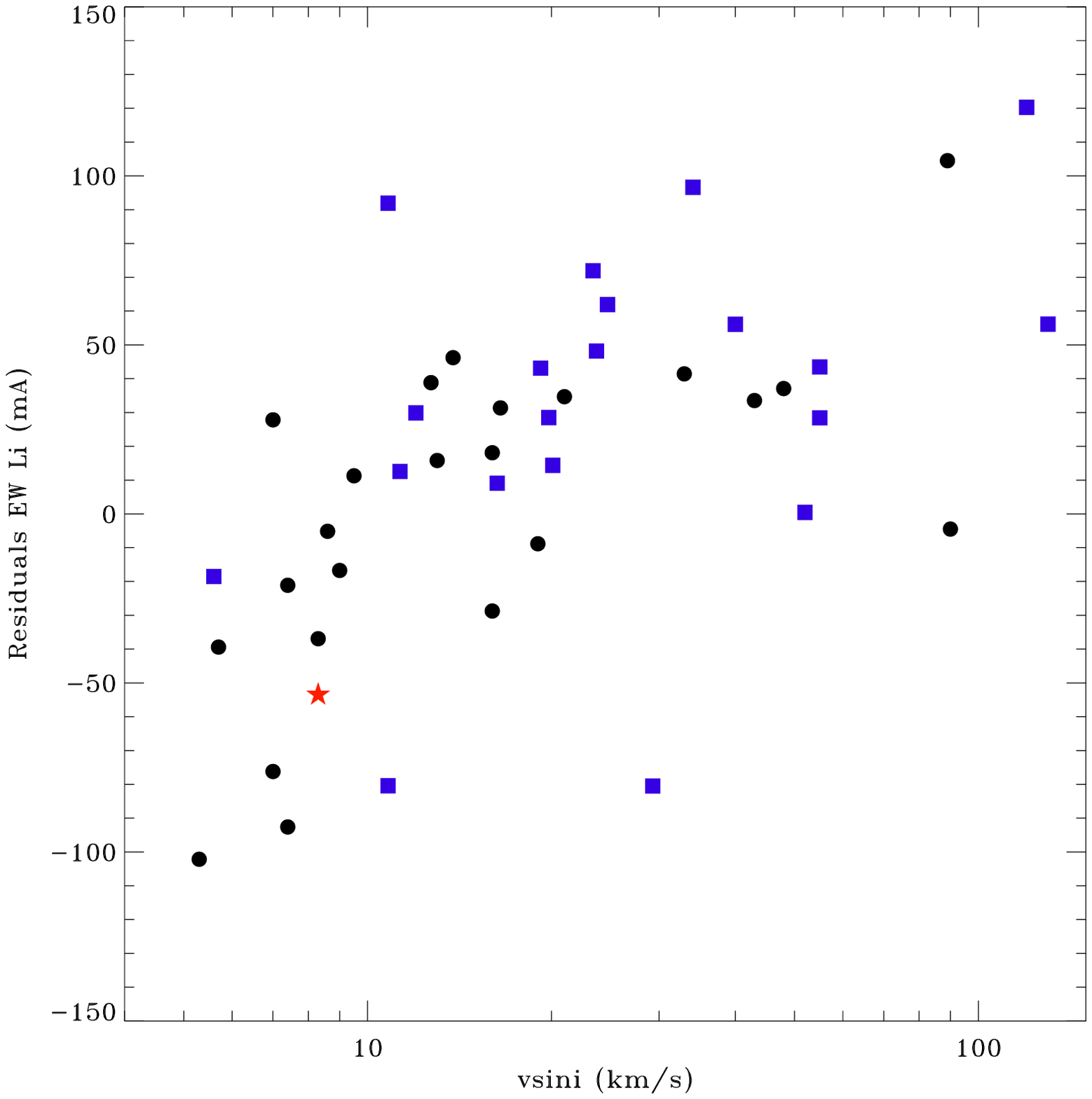}
   \caption{Upper panel: residuals of Li 6708 EW with respect to mean locus vs B-V defined by
   IC 2391 members plotted versus rotational period (dashed line in Fig~\ref{f:ewli}).
   Filled and open (black) circles: IC 2391 stars with confirmed+likely and uncertain rotation 
   periods, respectively;
   filled and open (blue) squares: Argus field stars,  with confirmed+likely and uncertain 
   rotation periods, respectively; 
   (red) star: HD\,61005. The continuous line 
   represents the fit of IC2391 members with confirmed rotational periods, and the dotted
   line represents the fit including the unconfirmed periods.
   Lower panel: residuals of Li 6708 EW with respect to mean locus vs B-V defined by
   IC 2391 members plotted versus projected rotational velocity. Filled circles: IC 2391. Filled squares:
   Argus field. (Red) star: HD\,61005. }
              \label{f:res3li}%
    \end{figure}

To take isuch a dependence on rotation into account, we derived the mean locus of IC2391 members
on the Li EW vs B-V diagram
and the residuals from it.
When plotting such residuals versus rotational period, a correlation is found with significant results
at about 2 $\sigma$ (Fig.~\ref{f:res3li}).
HD\,61005 fits such relation(s) fairly well, i.e. it has the expected lithium
content for IC2391 stars with its rotational period.
A similar trend is also seen considering the projected rotational velocity.

\subsection{Chromospheric and coronal activity}

\begin{figure}
   \centering
   \includegraphics[width=9cm]{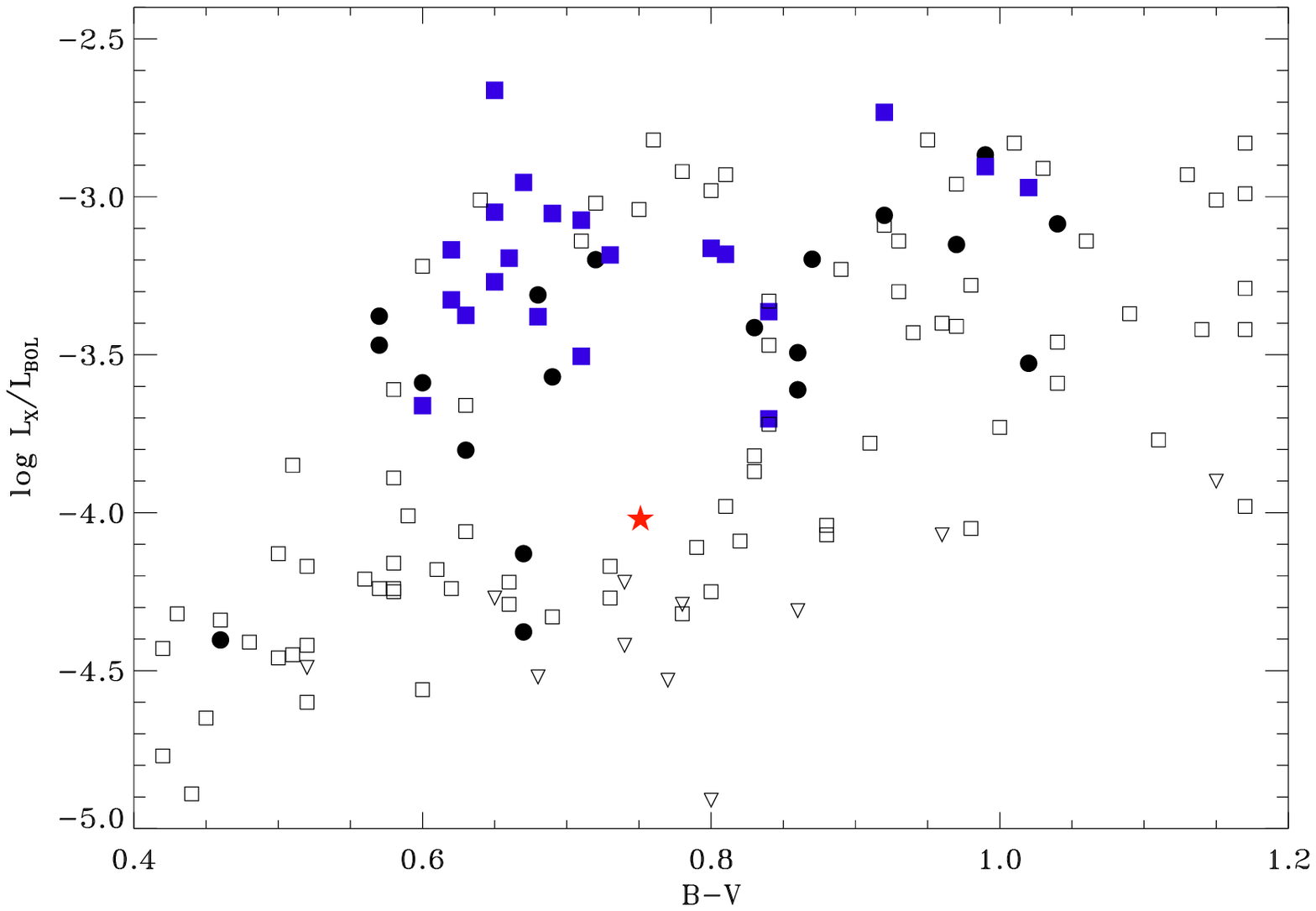}
   \caption{Distribution of $\log L_X/L_{bol}$ of IC 2391 (filled circles), 
            Argus association (blue filled squares), and
            Pleiades members (open squares and open triangles for detections and 
            upper limits respectively; data taken from
            Stauffer et al.~\cite{stauffer94})
            versus B$-$V colour. HD\,61005 is plotted as a (red) star. 
            A few IC 2391 members are not detected in X-ray. }
              \label{f:xray}%
    \end{figure}

There are no members of Argus association with measured H\&K emission, so no comparison is 
possible for this indicator. X-ray data of IC 2391 are taken from several sources and 
summarized in Appendix A. X-ray data of Argus field members 
are all taken from the ROSAT Bright Source Catalog.
Figure \ref{f:xray} shows the distribution of $\log L_X/L_{bol}$ for members of Argus association, IC 2391 open cluster, 
and HD\,61005.
It results that HD\,61005 has a lower X-ray luminosity than most of the Argus and IC 2391 members, where a few 
slow rotators have similar coronal luminosities. A few other members have no X-ray detections
in ROSAT All Sky Survey. 
A systematic difference between field stars and IC 2391 members is also seen  in X-ray emission, most likely
due to the selection effects previously discussed.

\subsection{Chemical composition}

\begin{figure}
   \centering
   \includegraphics[width=9cm]{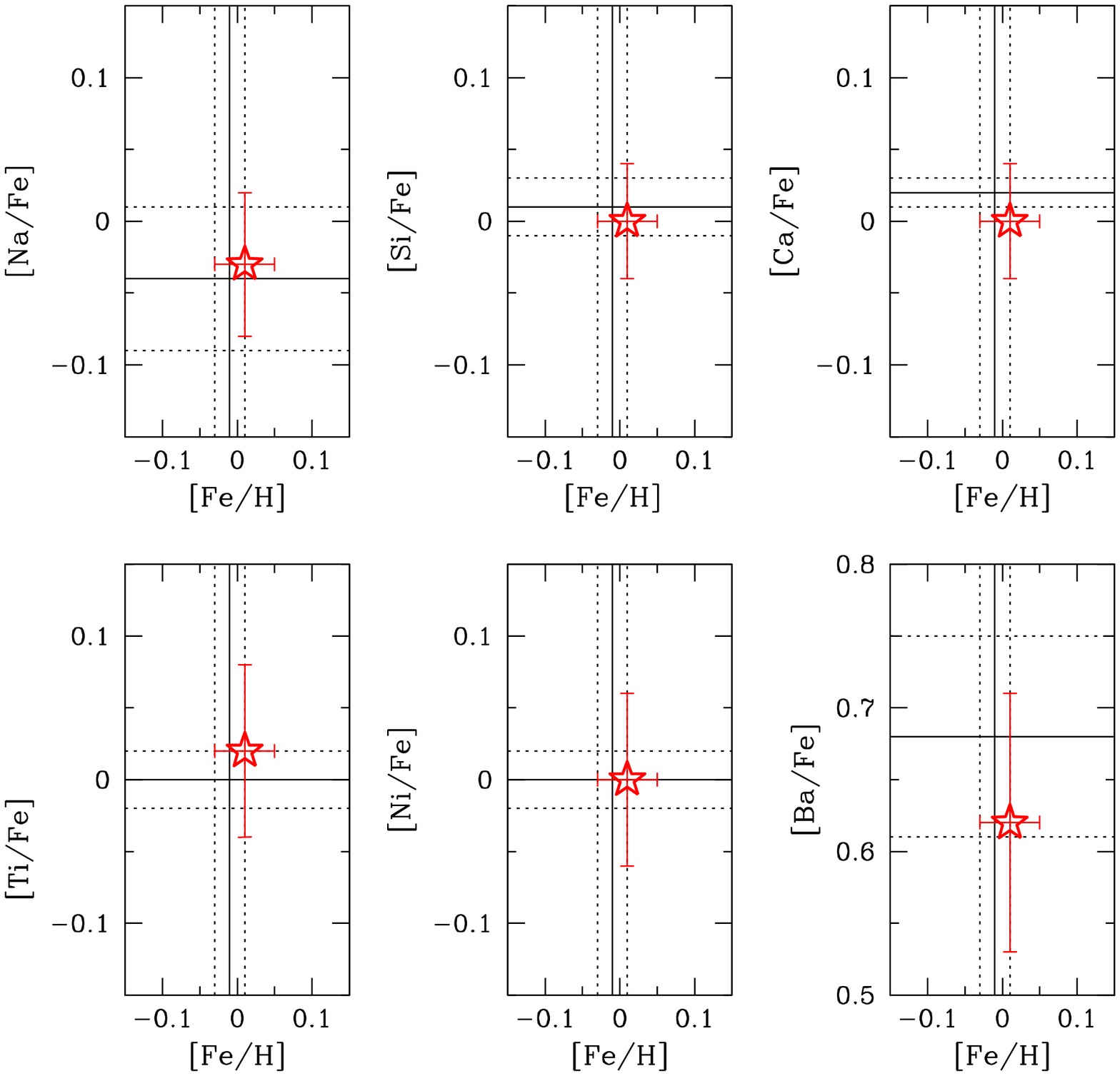}
   \caption{Abundances of individual elements of HD\,61005 compared to those of IC2391
  (from D'Orazi \& Randich \cite{dorazi09} and D'Orazi et al.~\cite{dorazi09a}).}
              \label{f:chem}%
    \end{figure}

Abundance ratios of iron-peak and $\alpha$-elements for HD~61005 are very similar to those derived for IC~2391 members by DR09,
reflecting in both cases a solar chemical composition. The mean difference in [X/Fe] ratios of our star with IC~2391
is only $\sim$ 0.02 dex, when considering all the
elements listed in Table~\ref{t:abu}. Our analysis was performed in the same way as that presented in 
DR09 and D'Orazi et al.~(\cite{dorazi09a}) for IC 2391 members, so that systematic errors in the abundance
scale are not of concern for such comparison.
In Fig.~\ref{f:chem}, we show [X/Fe] ratios as a function of [Fe/H]
for HD 61005  along with the average values  and their corresponding standard deviation
 as obtained from IC 2391 stars.
Similarly, Argus association field stars have a nearly solar iron content (Viana Almeida et al.~\cite{viana09}).
This provides further support for the link between HD 61005 and the Argus association.
However, concerning Fe, Na, Si, Ca, Ti, and Ni, IC~2391 does not have a peculiar abundance pattern with respect to
nearby young stars: all the young clusters/associations surveyed so far from a
chemical point of view show a composition that is very close-to solar (Santos et al.~\cite{santos08}, 
D'Orazi et al.~\cite{dorazi09b}, Biazzo et al.~\cite{biazzo10}).
This means that the chemistry cannot conclusively state the physical association of HD 61005 to Argus.

On the other hand, a very interesting datum is given by the Ba content: HD 61005 shares with IC 2391 dwarfs the
unusual, extremely high [Ba/Fe] ratio, which is about a factor of five above the solar value ([Ba/Fe]$\approx$0.7 dex).
As discussed in D'Orazi et al.~(\cite{dorazi09a}), this anomalous Ba abundance is likely due to NLTE effects
(probably arising from over-ionization phenomena caused by the high level of
chromopsheric activity that is typical of young solar type stars, see e.g. Schuler et al.~\cite{schuler10}) 
rather than to a real higher content of neutron-capture elements in young, nearby stars. 
However, our result points out that the [Ba/Fe] trend in HD 61005 is very different from the one derived for old, 
solar-type stars, providing an independent, complementary observational hint of the young nature of this
object (see D'Orazi et al.~\cite{dorazi09a} for a detailed discussion on the relationship between Ba and cluster age).

\subsection{IR excess}

As mentioned in Sect.~\ref{s:intro}, HD 61005 has a very large IR excess.
We briefly discuss this in the context of our revised age of the system, because
the occurrence of an IR excess and its amount with respect to the photosphere 
are both observed to decline with age. 
IC 2391 was observed at $24 \mu m$ with Spitzer by Siegler et al.~(\cite{siegler07}),
resulting in a frequency of FGK stars with IR excess of $31^{+13}_{-9} \%$, which fits
the trend toward decreasing IR excess as a function of age, as observed for 
other clusters.
No observations of IR excess have been published for the field members of the
Argus association.
HD 61005 has a $24~\mu m$ excess ($f_{excess}/f_{phot}=1.096\pm0.049$; Meyer et al.~\cite{meyer08})
that is more than for any star in IC 2391 included in the Siegler et al.~(\cite{siegler07}) sample, 
with the F3 star \object{HD74374}=PMM1174 just below this value ($f_{excess}/f_{phot}=0.98$).

No clear trend toward IR excess with stellar rotation results from 
IC 2391 data: IR excess is found on slow rotators like \object{PMM 4413}\footnote{PMM 4413 
is similar to HD 61005 in several aspects: it has a slightly earlier spectral type, slow rotation
($P_{rot}=5.05$~d), faint X ray emission ($ \log L_{X}/L_{bol}=-4.47$), Li EW below the mean
locus of the cluster and presence of $24~\mu m$ excess. However, it is a double-lined spectroscopic binary
with orbital period 90 days (Platais et al.~\cite{platais07}).}  and \object{PMM 4467}, but also
on the very fast rotator \object{PMM 1820}. Observations of the whole cluster, as well as the field
members of the association, might provide higher quality statistics for this kind of analysis.
A similar analysis for the Pleiades open cluster is also inconclusive: Greaves et al.~(\cite{greaves09})
note that the four solar type stars with $24~\mu m$ excess are slow rotators but selection effects
might be at work because half of fast rotators have no excess measurement, because they are projected 
in a region of heavy cirrus (Gorlova et al.~\cite{gorlova06}).

\section{Conclusion}
\label{s:conclusion}

We have examined several properties of HD\,61005 with the goals
of a better characterization of 
the system and a robust determination of its age.
We determined rotation period, radial and projected rotational velocity,
chromospheric emission, effective temperature, and chemical composition
(see Table \ref{t:summary} for a summary of target properties).
From projected rotational velocity and rotation period we inferred that the
star is seen close to edge-on. In that case, there is no indication of any significant
misalignment with the debris disk. 

\begin{table}[h]
\caption{Summary of parameters of HD~61005} 
\label{t:summary}
\begin{center}       
\begin{tabular}{lcl} 
\hline
Parameter       &  Value           & Reference  \\
\hline
V                      & 8.22$\pm$0.01    & this paper \\
B-V                    & 0.751$\pm$0.003  & Menzies et al.~(\cite{menzies90}) \\
V-I                    & 0.805$\pm$0.005  & Menzies et al.~(\cite{menzies90}) \\
$V-K_{S}$              & 1.762$\pm$0.027  & this paper \\
J-K                    & 0.447$\pm$0.036  & 2MASS \\
$\pi$ (mas)            & 28.29$\pm$0.85   & van Leuveen (\cite{vanlee07})   \\
$M_{V}$                & 5.478$\pm$0.068  & this paper \\
$L/L_{\odot}$          & 0.583$\pm$0.048  & this paper \\
$R/R_{\odot}$          & 0.840$\pm$0.038  & this paper \\
$T_{eff}$              & 5500$\pm$50      & this paper \\
$ \log g $             & 4.5$\pm$0.2      & this paper \\
${\rm [Fe/H]}$         & 0.01$\pm$0.04    & this paper \\
EW Li 6708 (m\AA)      & 171$\pm$3        & this paper \\
A(Li)                  &  2.85$\pm$0.1    & this paper \\
S index                &  0.502           & this paper \\
$\log R_{HK}$          & -4.310           & this paper \\
$ log L_{X}$           & 29.3$\pm$0.2     & this paper \\
$ log L_{X}/L_{bol}$   & -4.02            & this paper \\
$ v \sin i$ (km/s)     & 8.2$\pm$0.5      & this paper \\
$P_{rot}$ (d)          & 5.04$\pm$0.04    & this paper \\
$P_{cyc}$ (y)          & 3.4$\pm$0.1      & this paper \\
$ i_{star}$            & 77$^{+13}_{-15}$ & this paper \\  
RV (km/s)              & 22.5$\pm$0.5     & this paper \\
$\mu_{alpha}$ (mas/yr) & -55.71$\pm$0.59  & van Leuveen (\cite{vanlee07}) \\
$\mu_{delta}$ (mas/yr) &  74.58$\pm$0.62  & van Leuveen (\cite{vanlee07}) \\
U (km/s)               & -22.2$\pm$0.6    & this paper \\
V (km/s)               & -14.3$\pm$0.3    & this paper \\
W (km/s)               & -4.1$\pm$0.2     & this paper \\
age (Myr)              &  40              & this paper \\
\hline
\end{tabular}
\end{center}
\end{table}

We found that when applying the standard age calibrations to the values
of several age indicators (lithium, chromospheric and coronal emissions,
rotation period) HD\,61005 results of age comparable to that of the Pleiades
(120 Myr).
However, the kinematic parameters strongly indicate membership in the 
Argus association, which is significantly younger (40 Myr; Torres et al.~\cite{torres08}).
We then compared the properties of HD\,61005 to those of Argus association members,
including the open cluster IC 2391.
We found that HD\,61005 parameters are on the edges but not outside the distribution of IC2391 members.
The lithium content and coronal emission are similar to those of 
members of IC 2391 of comparable rotational period.
HD\,61005 also has a similar chemical composition to IC 2391.

HD\,61005, therefore, can have the same age as the slowly rotating, less active, 
and (relatively) Li-poor stars in IC 2391. The kinematic parameters
strongly indicate that this is indeed the case.
We therefore conclude that the association with Argus is very likely and the age 
of HD\,61005 is about 40 Myr.

The younger age for HD\,61005 than currently assumed (90 Myr, Hines et al.~\cite{hines07})
has a significant impact on the detection limits of planetary companions in direct imaging
programs.
Buenzli et al.~(\cite{buenzli10}) derive limits of about $3$ and $5~M_{J}$ at 50 and 15 AU, respectively, 
assuming a 90 Myr age.
These limits become significantly smaller when taking the revised age into account.
Therefore, only less massive planets may perturb the disk and be a likely
cause for the peculiar features described in Sect.~\ref{s:intro}. 
Dedicated modeling of the possible masses and locations of planets that might explain the
disk features are needed. 

A younger age also helps in explaining the unusually large IR excess observed for HD\,61005,
since debris disk frequency and luminosity are known to decline with age (Evans~\cite{evans09}; Wyatt~\cite{wyatt08}).
However, even at an age of 40 Myr, the $24~\mu m$ excess of HD\,61005 still turns out to be larger
than that of any IC 2391 member for which suitable data are available. 
The solar abundance of HD\,61005 is consistent
with the lack of significant metallicity differences between stars with debris disks and 
general field stars, at odds with what has been found for the parent stars of giant planets
(Greaves et al.~\cite{greaves06}).

Finally, from our comparison of HD\,61005 with the Argus association, differences 
can be noted between field and cluster members in  Li EW, X-ray luminosity, and possibly 
in rotational period distributions.
Argus field stars are on average more active, rotate faster, and have more lithium 
than their IC 2391 cluster counterparts.

A small age difference (5-10 Myr), with  field stars being slightly younger than
cluster members, might explain the lithium difference.
A vigorous star formation event followed by supernova explosion(s)
might have triggered star formation in nearby regions, with some delay after
the cluster was formed. 
This kind of scenario is envisaged for the Sco-Cen complex as a sequence of star
formation events, involving \object{Lower Centaurus Crux}, \object{Upper Centaurus Lupus}, 
\object{Upper Scorpius}, 
\object{TW Hyd association},  \object{$\epsilon$} and \object{$\eta$ Cha} 
(Ortega et al.~\cite{ortega09}). 

However, our favorite explanation for these differences is a selection effect of Argus field stars 
all being selected from the SACY sample (Torres et al.~\cite{torres06}), which only includes stars with X-ray counterparts 
in the ROSAT Bright Star Catalog (Voges et al.~\cite{rosatbright}), while IC\,2391 members are selected using 
a variety of membership indicators.
Our adopted target list of IC 2391 members from Torres et al.~(\cite{torres08}) includes additional
selection criteria aimed at better defining the kinematical properties of the association
(excluding known or suspected spectroscopic binaries unless an orbital solution is available). 

Our results suggest that, at least for the associations beyond 100 pc in the
Torres et al.~(\cite{torres08}) list, a selection criterion based on inclusion in the ROSAT Bright Star Catalog
might have caused a bias towards the most active and fast rotating stars. 
Therefore, we suggest that other less active, slowly rotating members probably still wait to be identified.

If the different disk lifetime is one of the causes of the dispersion of stellar rotation rates in coeval groups
of young stars  and if planet formation occurs more easily in long-lived disks, the selection bias 
that we identified might have a significant impact for disk and planet searches.
The correlation of lithium and other age indicators with stellar rotation might also
cause the age determination for isolated field stars to be biased if such a dependency is 
not taken into account.

\begin{acknowledgements}
We thank the NaCo Large Program Collaboration for Giant 
Planet Imaging (ESO program 184.C-0567) for calling our attention to HD\,61005.
We thank R. Wichmann and J.H.M.M. Schmitt for useful information on the
X-ray properties of the star, for checking the original RASS image,
and for providing an updated value of the X-ray count rate.
This research has made use of the SIMBAD database, operated at CDS, Strasbourg, France
J. C. C. was supported in part by NSF AST 1009203.

\end{acknowledgements}

{}

\appendix
\section{X-ray emission of IC 2391 members}
\label{s:xray_ic2391}
Because the IC2391 cluster extends over an area
larger than the one considered in pointed X-ray observations and because several new members
are considered, we summarize here the X-ray luminosities of the IC2391 members 
selected from Torres et al.~(\cite{torres08}).
We collected the results of pointed observations with ROSAT (Patten \& Simon \cite{patten96};
Simon \& Patten \cite{simon98}) and XMM (Marino et al.~\cite{marino05}).
We also cross-checked the Torres et al.~(\cite{torres08}) membership list with the
ROSAT all-sky bright and faint sources (Voges et al.~\cite{rosatbright}; \cite{rosatfaint}), 
adopting a matching radius of 30 arcsec.
X-ray fluxes were derived using the calibration by H\"unsch et al.~(\cite{hunsch})
and X-ray luminosities adopting the distances from Torres et al.~(\cite{torres08}).

\begin{table}
   \caption[]{X-ray emission of IC 2391 members.  }
     \label{t:x_ic2391}
      
       \begin{tabular}{rcccccc}
         \hline
         \noalign{\smallskip}
         Star & Star & 1RXS & V & B-V & $\log L_{X}$ & Ref. \\
         PMM  & VXR  &      &   &     &              &             \\ 
        \noalign{\smallskip}
         \hline
         \noalign{\smallskip}

 7422 &      &   J082845.5-520523 &  10.40 &  0.69 &             29.95  &              1 \\	
 7956 &      &   J082952.7-514030 &  11.50 &  0.97 &             30.17  &              1 \\ 
 6978 &      &   J083502.7-521339 &  12.01 &  1.02 &             29.52  &              2 \\
 2456 &      &   J083543.4-532123 &  12.16 &  0.92 &             29.94  &              2 \\
  351 &      &   J083624.5-540101 &  10.17 &  0.57 &             30.26  &              1 \\
 4336 & 2b   &   J083756.4-525714 &  11.25 &  0.86 &             29.98  &              2 \\
      &      &                    &        &       &             29.72  &              3 \\   
 4362 & 3a   &                    &  10.91 &  0.67 &             29.36  &              3  \\
 4413 & 5    &                    &  10.20 &  0.67 &             29.36  &              3 \\
 4467 & 12   &   J083954.3-525755 &  11.80 &  0.86 &             29.52  &              2  \\
      &      &                    &        &       &             29.63  &              3 \\
      &      &                    &        &       &             29.39  &              4 \\
 1083 & 14   &   J084007.4-533744 &  10.38 &  0.57 &             30.15  &              1  \\
      &      &                    &        &       &             30.17  &              3  \\ 
 8415 & 16a  &   J084016.6-525629 &  11.63 &  0.87 &             30.03  &              2 \\
      &      &                    &        &       &             30.03  &              3 \\ 
      &      &                    &        &       &             30.04  &              4  \\
 1759 & 18   &                    &  12.54 &  1.25 &             29.59  &              3 \\
 1142 & 22a  &   J084049.6-533737 &  11.04 &  0.68 &             29.93  &              2 \\
      &      &                    &        &       &             30.26  &              3 \\   
 1820 & 35a  &   J084124.0-532247 &  12.41 &  0.99 &             30.22  &              1  \\
      &      &                    &        &       &             30.11  &              3  \\ 
 4636 & 41   &                    &  13.20 &  1.36 &             29.66  &              3   \\
      &      &                    &        &       &             29.75  &              4   \\
      &      &                    &        &       &             29.88  &              5  \\
 3695 & 47   &                    &  13.30 &  1.44 &             29.57  &              3  \\
      &      &                    &        &       &             29.79  &              4   \\
      &      &                    &        &       &             29.72  &              5  \\
 2888 & 66   &                    &   9.76 &  0.46 &             29.43  &              3 \\
 2012 & 69a  &                    &  11.41 &  0.83 &             29.85  &              3 \\
 4809 & 70   &  J084407.4-525316  &  10.73 &  0.63 &             29.73  &              2  \\
      &      &                    &        &       &             29.84  &              3  \\  
 5884 & 72   &  J084425.2-524219  &  11.35 &  0.72 &             29.96  &              2   \\
      &      &                    &        &       &             30.18  &              3  \\    
 4902 & 76a  &  J084526.6-525132  &  12.45 &  1.04 &             29.70  &              2   \\
      &      &                    &        &       &             29.93  &              3  \\  
 6811 & 77a  &  J084539.5-522556  &   9.91 &  0.60 &             30.24  &              1  \\
      &      &                    &        &       &             30.18  &              3  \\   
\hline
\multicolumn{7}{c}{X-ray non detections} \\
\hline
 1560 &      &   &  10.64 &  0.61 &               $<$  &             \\ 
 6974 &      &   &  12.04 &  1.04 &               $<$  &             \\ 
 4280 &      &   &  10.04 &  0.67 &               $<$  &             \\ 
 3359 &      &   &  11.38 &  0.76 &               $<$  &             \\ 
 5376 &      &   &  13.94 &  1.37 &               $<$  &             \\ 
  665 &      &   &  11.33 &  0.75 &               $<$  &             \\ 
  686 &      &   &  12.55 &  1.04 &               $<$  &             \\ 
  756 &      &   &  11.06 &  0.68 &               $<$  &             \\ 
 5811 &      &   &   9.16 &  0.37 &               $<$  &             \\ 
 1373 &      &   &  12.06 &  0.96 &               $<$  &             \\ 
 2182 &      &   &  10.18 &  0.63 &               $<$  &             \\ 

         \noalign{\smallskip}
         \hline
      \end{tabular}

Sources of data: 1: ROSAT Bright Source Catalog
(Voges et al.~\cite{rosatbright}); 2:  ROSAT Faint Source Catalog 
(Voges et al.~\cite{rosatfaint}); 3:  Patten \& Simon (\cite{patten96});  
4: Simon \& Patten (\cite{simon98}); 5: Marino et al.~(\cite{marino05}).

\end{table}

\end{document}